# Fundamental limitations of Huygens' metasurfaces for optical beam shaping

Carlo Gigli[1], Qitong Li[2], Pierre Chavel[3], Giuseppe Leo[1], Mark Brongersma[2], Philippe Lalanne[4*]

[1] Matériaux et Phénomènes Quantiques, Université de Paris & CNRS, 10 rue A. Domon et L. Duquet, 75013 Paris, France
[2] Department of Materials Science and Engineering, and Geballe Laboratory for Advanced Materials, Stanford University, Stanford, California 94305, USA
[3] Laboratoire Hubert Curien, Université de Lyon, Univ. Jean Monnet de Saint Etienne, Institut d'Optique Graduate School, CNRS UMR 5516, 42000 Saint Etienne, France
[4] Laboratoire Photonique, Numérique et Nanosciences (LP2N), Institut d'Optique Graduate School, Univ. Bordeaux, CNRS, 33400 Talence Cedex, France.
[*]Corresponding author: Philippe.lalanne@institutoptique.fr

Optical dielectric metasurfaces composed of arrayed nanostructures are expected to enable arbitrary spatial control of incident wavefronts with *subwavelength spatial resolution*. For phase modulation, one often resorts to two physical effects to implement a $2\pi$-phase excursion. The first effect relies on guidance by tall nanoscale pillars and the second one exploits resonant confinement by nanoresonators with two degenerate Mie-resonances. The first approach requires high aspect ratios, while the second one, known as Huygens' metasurfaces, is much flatter, and thus easier to manufacture. We compare the two approaches, more focusing on conceptual rather than technological issues, and identify fundamental limitations with the latter. We explain the origin of the limitations based on general arguments, such as reciprocity, multimode/monomode operation and symmetry breaking.

## 1. Introduction

In integrated electronics, an important prerequisite for success is that every transistor or memory element can be driven or read independently, and nanotechnology is used to push the integration limit subject to that essential requirement. In dielectric metasurfaces for wavefront shaping, one aims at imprinting spatially varying phases (or amplitudes, polarizations) on an incoming optical beam, and nanotechnology is used to spatially control the phase down to the subwavelength scale, as required for high-numerical-aperture focusing or imaging, beam steering over large angles, and computer-generated holography. Just like for integrated circuits, it is important that the phase (or amplitude, polarization) can independently be controlled by adjacent nanostructures (or meta-atoms) at the sub-wavelength scale. This demanding requirement has marked the history of diffractive optics and has led to the emergence of two families of metasurfaces associated to different mechanisms: guidance along nanoscale pillars, implementing wavefront phase differences [1]; and nanoconfinement with two overlapped resonances, each covering a standard phase range of $\pi$ [2–4].

Many good reviews exist on the topic [5–10], and exactly the same mechanisms have also been used to control the geometrical phase [11–13]. Geometric-phase control will not be analyzed hereafter; however, we expect that the following conclusions remain qualitatively valid.

Guidance along nanopillars has long been known for providing high diffraction efficiencies (~80% routinely) at large deviation angles [1,13,14] and being resilient to changes of the illumination direction[15]. It is intrinsically a non-resonant mechanism, which is presently attracting renewed interest for various applications [8,9,11,13,16–18]. The more recent nanoconfinement approach [2–4,11,19] is well understood from a simple picture where every meta-atom provides two electric and magnetic dipolar resonances [2,20] with opposite phases and achieve forward scattering [21]. Together, these resonances make every meta-atom behave as a secondary source emitting purely forward-propagating elementary waves, hence the name of Huygens' metasurfaces[2] that is often given to this resonant approach.

Importantly, the resonant approach requires much smaller aspect ratios than the guidance one, and thus is more readily amenable to low-cost large-scale manufacturing. Owing to this promise and its reputation of excellent performance, it is drawing a strong attention [22–27].

In this work, we question the applicability of the resonant approach for beam shaping. More generally, we highlight the lack of optical meta-atoms capable to implement advanced functionalities, e.g. forward scattering, and quasi-independent operation with a weak crosstalk when arrayed with a sub-wavelength or even wavelength-scale distance. The issue is not anecdotal and may be even seen as a fundamental limitation, as will be shown at the end of the article by making a parallel with previous works on optical metamaterials.

Resonant optical metasurfaces have found plenty of applications[5] in the last decade and will open new challenges and opportunities, e.g. for sensors or active metasurfaces[28]. While we feel that beam shaping is limited by the limitations discussed here, some other applications may be affected more marginally. We therefore stress that the present analysis is not intended to undervalue any previous work. It is rather expected to offer a helpful view which is, to our knowledge, missing in the literature, so as to call attention to limitations and, hopefully, contribute to the development of the field from a different perspective. Some aspects touched upon here are already well-known to readers with different backgrounds. However, we think that, as a community, we have sometimes underappreciated the importance of these limitations and overlooked their fundamental character.

In a first step, in Section 2, we provide a context to the problem by studying the efficiency of a series of meta-gratings with increasing periods, targeting applications in the visible and near-infrared. To illustrate our purpose, we arbitrarily chose the pioneering Huygens' metasurface design made of arrayed silicon nanodisks [2], but similar conclusions remain qualitatively valid for more recent designs. We need a benchmark for assessing the performance and the origin of the limitations. Thus, comparison with the nanopillar guidance approach will be highlighted throughout the article. Our numerical calculations show that the efficiency of the Huygens' elements remains below 40% for monochromatic operation, even for meta-gratings with periods significantly larger than the wavelength for which classical low-cost sawtooth diffractive optical elements show excellent performance, around 90%.

This low performance is known in the field of Huygens's metasurfaces and ascribed to interelement electromagnetic coupling [9,25,29]. In most recent works, it is stated as a contingent constraint to be overcome with more clever designs or numerical optimization. Section 3 carefully reviews recent efforts. In particular we recall the attempts to improve Huygens' metasurface performance through electromagnetic computations, essentially by local periodization[29,30] or global optimization[25,27] of metaatoms features, sizes and positions. Significant improvements have been reported theoretically in some cases, but the experimental results remain inferior to those of sawtooth diffractive optical elements. We emphasize that a global optimization, at the scale of the entire period, synchronizes the scattering of an ensemble of elements towards a specific goal. In no way, it may result into an individual control of the phase at the subwavelength scale. The reverse holds true actually.

Although these findings suggest that the promised benefits of nanotechnology are not fully gathered in Huygens' metasurface, they are so-far based on observations made for specific meta-atoms (the reference Huygens' design analyzed in Section 2 or optimizations performed by other researchers summarized in Section 3). In Section 4, we question the applicability of resonant optics for high numerical aperture focusing or imaging or beam shaping, based on basic concepts, e.g. cross-sections, crosstalk, number of modes, symmetry breaking, reciprocity. We thereby hope to reach comprehensive conclusions about the fundamental limitations announced in the title.

We emphasize again that our discussion holds for *beam-shaping applications requiring a subwavelength control of some wave features* (phase, amplitude, polarization) and for operation in the visible domain with transmit arrays and dielectric materials. At longer wavelengths, where materials and technologies are different, our general statements are still worth considering in our opinion, although different practical conclusions might apply. We also expect that our arguments may help the design of plasmonic metasurfaces. Note that our conclusions do not hold for reflect arrays, see SubSection 4.3.

**2. Numerical evidence for a testbed case**

Let us consider the initial Huygens' design reported in [2], based on Si nanodisks embedded in a homogenous medium ($n_h = 1.66$) (Figure 1b). We also consider TiO$_2$ square-based nanoposts on a SiO$_2$ substrate (Figure 1a) [14].

We do not intend to provide a thorough comparison between the two approaches. The latter rely on different materials and operating wavelengths. A single target response is considered; it would be relevant to analyze different targets, e.g. the frequency dispersion or possible multi-color operation. In addition, the guidance approach requires large aspect ratios, and is thus inconvenient for mass production. Conversely, the resonance approach is quite sensitive to tiny imperfection and requires a precise control of meta-atom sizes. In fact, the guidance approach, which is well documented [1,7,14,15], is used as a reference point to better exhibit the limitations of the resonant approach and to feed the reflection with quite different perspectives.

The initial design of any metasurface often relies on the creation of a lookup table that provides a systematic, one-to-one mapping between the desired phase profile and the required meta-atoms [2,14]. Lookup tables can be computed very quickly and offer an initial guess for the ideal choice of meta-atom, which may then facilitate rapid prototyping or further computationally demanding global optimizations. They are usually computed by considering periodic arrays of identical meta-atoms illuminated at normal incidence and extracting the phase of the transmitted zeroth-order as a function of some tuning parameters, in general the meta-atom size or shape. Let us note that this natural approach is optimal at least for metasurfaces encoding phase increments significantly smaller than $2\pi/\lambda$ between two successive elements. However, for rapidly varying phases, the neighboring meta-atoms of functional metasurfaces are all quite different and the lookup table based on an artificial periodization ceases to be strictly valid. In addition, the computation being performed at normal incidence, the lookup table includes some symmetries that are lost in the functional metasurfaces. These remarks will help us in the following to understand why lookup tables are not always helpful to evaluate the performance of functional – in particular Huygens' – metasurfaces.

The lookup tables obtained for normal incidence are reported in Figure 1. They are consistent with previous data in [14] and [2], evidencing a near-unity transmission for the required $2\pi$ phase excursion. Please note that due to the symmetry of the systems these results are valid for both polarizations along the $x$ and $y$ axes.

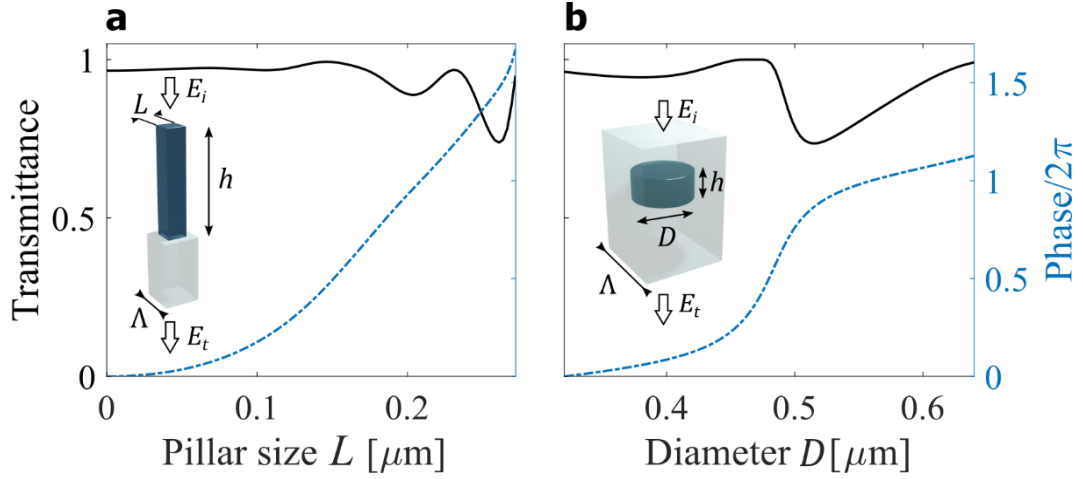

Figure 1. Lookup table of the zero-order transmission coefficient of gratings formed by identical meta-atoms organized in a square lattice of period $\Lambda$: modulus (black solid line) and phase (blue dash-dot line). **a** Geometry in [14], with TiO$_2$ ($n_{TiO_2} = 2.3$) square-basis nanopillars on a SiO$_2$ substrate ($n_s = 1.46$), $\Lambda = 272$ nm and $\lambda = 633$ nm. **b** Geometry in [2], with Si ($n_{Si} = 3.5$) nanodisks embedded in a host medium ($n_h = 1.66$), $\Lambda = 666$ nm and $\lambda = 1340$ nm. A plane wave normally incident on the grating from above and linearly polarized is considered in both cases. The numerical results in **a** are obtained with the rigorous coupled-wave analysis (RCWA) implemented with the RETICOLO freeware [31]. Those in **b** are obtained with the RCWA solver [32] for circular structures and with COMSOL Multiphysics.

The lookup tables are then used to design two series of blazed gratings with increasing periods, each designed to produce the same sawtooth phase profile. The gratings with the smallest period $a = 3\Lambda$ contain three meta-atoms. The periods are then increased step by step up to $a = 10\Lambda$. The design procedure is exactly the same for the two platforms. For metasurfaces with $N$ meta-atoms per period, the disk diameters and pillar sizes are chosen from the lookup tables to implement a phase profile $\varphi(x_p) = \frac{\pi}{N} + \frac{p2\pi}{N}$ with $x_p$ the position of $p^{th}$ element and $p \in [0,1,...N-1]$. The procedure is detailed in [14].

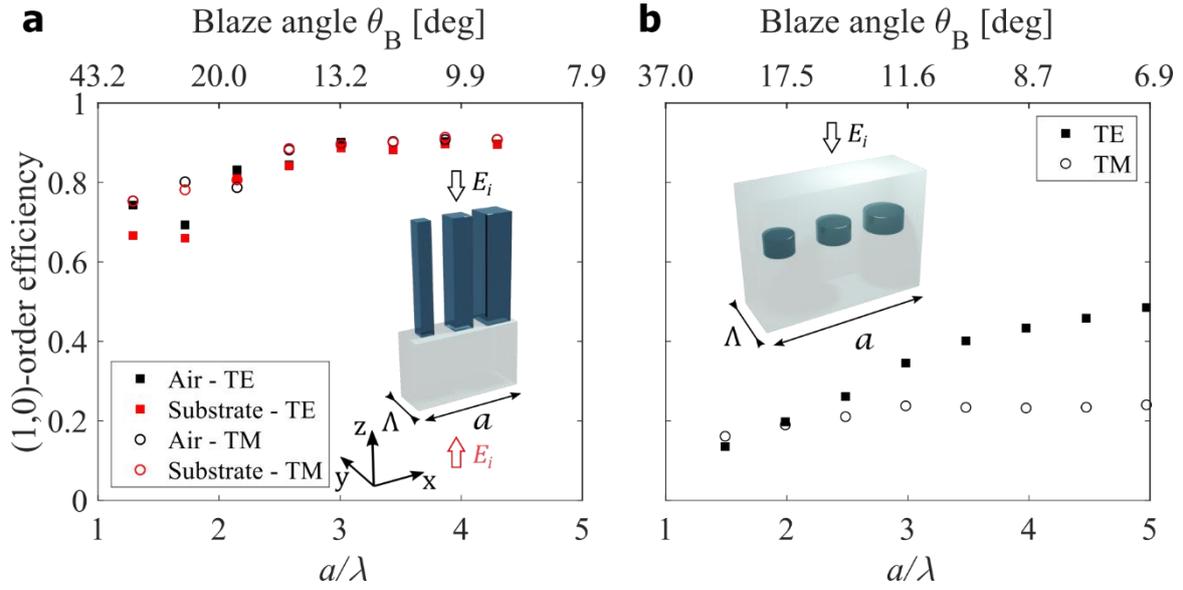

Figure 2. Comparison between the theoretical performances of meta-gratings designed with the lookup tables described in Figure 1 to be blazed into the (1,0) order. The unit cell dimensions of the meta-gratings are $N\Lambda = a$ along the $x$-axis and $\Lambda$ along the $y$-axis, where $N$ is the number of meta-atoms in the unit cell. **a** Constitutive meta-atoms are $TiO_2$ nanopillars with square basis ($\Lambda = 272$ nm, $\lambda = 633$ nm). **b** Constitutive meta-atoms are Si nanodisks ($\Lambda = 666$ nm, $\lambda = 1340$ nm). The results are obtained for plane waves, polarized along either $y$ (TE) or $x$ (TM) and normally incident from either the substrate or the cladding (the four cases are represented with different colors or signs). Due to the system symmetry substrate and cladding incidence results are identical in **b**. The number of Fourier components retained in the RCWA computation is $(17N \times 17)$ along the $(x, y)$ directions. In **a**, the blaze angle is given in the substrate.

The results obtained in Figure 2 for the $TiO_2$ meta-gratings are consistent with earlier works [1,14]. At large periods, the first-order efficiency reaches a plateau; all the transmitted light feeds the first order, the missing energy being reflected due to an impedance mismatch. The efficiency decrease for small period-to-wavelength ratios mainly results from an under-sampling of the phase, only three waveguiding nanoposts being used for the smallest period. Decreasing the lattice of period $\Lambda$ does to include more nanopillars per supercell does not increase the efficiency, but slightly higher efficiency is obtained by optimizing the pillar sizes and relative positions.

In sharp contrast, a much weaker performance is obtained with Huygens' nanodisks. The missing energy feeds all transmitted and reflected orders, therein implying that neither a phase gradient nor the Kerker condition are achieved, see Supplementary Section S1 for more details. The weak efficiencies reported in Figure 2b are consistent with available experimental results based on lookup table designs. In [4], meta-atoms etched in polycrystalline silicon are used to steer by 14° a laser beam at 1550 nm with a measured efficiency of 36%. In [3], an efficiency of 45% is measured for a meta-grating that diffracts a laser beam at 705 nm by 10°. Similar performance is reported in [26] for a meta-grating operating at 650 nm with 36% efficiency.

Thus, the initial design proposed in [2–4] with Si nanodisks does not meet the expectation in terms of efficiency, even for low deviation angles for which conventional sawtooth diffractive elements provide large efficiencies at lower fabrication costs. We have verified, with numerical results not reported here, that similar conclusions also hold for more elaborate Huygens' metasurfaces recently designed with quadrupoles (instead of dipoles) [22].

We conclude that the lookup table computed for uniform arrays of identical meta-atoms are, in general, not reliable indicators for the performance of functionalized Huygens' metasurfaces implementing graded variations, even when the spatial gradients are small compared to the wavenumber. Several causes will be found in Section 4.

## 3. Optimization of Huygens' metasurfaces: an overview of recent results

Although the issue is not highlighted as straightforwardly as in Fig. 2b, the lack of agreement between lookup table expectations and actual performance has been mentioned in the literature [3,9,25,27,29]. The problem arises from the strong variations in electromagnetic-coupling which occur between elements of the unit cell and are not accounted for in the lookup table based on translation invariance. In this Section, we thoroughly review the efforts made to compensate for these variations and enhance the efficiency.

In [29], the authors propose to mitigate the impact of the coupling variation between nearest neighbors with a supercell approach based on the repetition of four identical Si nanodisk metaatoms. The approach that clearly reduces the possibility to control the phase with a subwavelength resolution has been tested for a hologram operating at 785 nm; it does not seem to markedly enhance the performance with a measured efficiency of 24%.

Instead of trying to avoid or minimize the interactions among meta-atoms, in [30], the authors spatially vary the lattice period to modify the interelement coupling and consequently the phase of the transmitted electric field. However, this approach results in large supercells (more than ten times the operating wavelength) which locally satisfy the translational invariance, thus preventing any sub-λ phase sampling and control.

Alternatively, one may resort to global optimization by locally adjusting the metaatoms morphology initially designed with the lookup table. A few examples are reported in the literature. In [25], the performance of a meta-grating has been optimized by slightly changing the size and position of the nine Si nanodisks arrayed to form the grating period. An experimental efficiency of 57% in the first order (10° deviation) has been measured for normal incidence and for the designed linear polarization at 1340 nm (a large efficiency drop is reported for the other polarization). Another example is found in [27], in which global optimization of hundreds of $TiO_2$ nanodisks has been performed with a genetic algorithm and cylindrical lenses of a few tens of microns in diameters were fabricated with a high-precision fabrication process for green light operation. Compared to conventional designs based on lookup table, significant performance improvement has been demonstrated. For instance, a 24-μm-diameter lens with a 60% efficiency and a 0.51 NA has been manufactured. Note that this interesting result holds for normally-incident light and linearly polarized light like in the previous example. Further note that cylindrical lenses, like meta-gratings, are translation variant only in one direction and that a full control of the coupling variations is not required.

These performances, albeit largely improved compared to those achieved with classical lookup table design, remain quite lower than those achieved with the nanowaveguide approach for similar devices, see [1,8,9,11,13,14,16–18,33]. They are also lower than the performance achieved with optimized multilevel profiles, see for instance [34,35]. Inverse design with optimization always increases the performance, especially for gratings. However, the approach requires a good initial guess and is computationally intense, especially for non-periodic metasurfaces performing functions like hologram generation. When global optimization results in a strong performance increase, the result is often highly angle and wavelength sensitive [25,27]. This is a consequence of the non-local behavior of the optimized geometry. Overall, we emphasize that global optimization contributes by no means to a subwavelength scale control of the phase. On the contrary, it favors a macroscopic collective behavior. Its philosophy is therefore at variance from a nice conceptual paradigm that leads to meta-atoms operating independently like in integrated circuits.

In our opinion, the substantial role played by non-local interactions in Huygens' metasurface performance deserves a thorough analysis and the question arises whether these interactions result in fundamental limitations restricting the metasurface capabilities for beam shaping.

## 4. Explanation for the low performance observed for the testbed case

As we have seen, the design with lookup tables has a limited scope and refinements through global optimization do not fundamentally change the situation. Let us now deepen our insight in the consequences of non-local interactions. Being again helped by the benchmark of the guidance approach, we expect to further appreciate whether better performance may be achieved with smart designs.

### 4.1. Shadowing-like effect at phase jumps

Figure 3 compares the near-field maps of nanodisks with different diameters, all for the same wavelength, TE-polarization and incident angle $\theta_i = 0°$. Sixteen nanodisks are considered. They are isolated in Fig. 3(a), periodically arrayed in (b) exactly as for the computation of the lookup table of Fig. 1(b), and arrayed to form a meta-grating in Fig. 3(c). The diameters are chosen to uniformly sample the $[0,2\pi]$ phase interval. The total period $a \approx 10.66$ µm of the meta-gratings corresponds to a first-order deviation angle of 4.3°; the meta-grating efficiency is 60% for TE-polarization.

All the maps are plotted with the same color scale for the sake of comparison. As evidenced by a direct visual inspection, the near-field maps significantly differ, therein again evidencing that the arrayed nanodisks are electromagnetically coupled and do not behave as independent entities. Focusing our attention on (b) and (c), we note that the most spectacular difference occurs on the right and left sides of the $2\pi$ phase jump location. The jump forces the association of small and large nanodisks and thus strongly breaks the translational invariance used for the lookup table computation. That explains why the lookup-table field maps are not recovered for the nanodisks neighboring the jump, in a $\approx 4\lambda$-wide "transient zone" corresponding to a total of about 8 pillars (4 on each side of the jump).

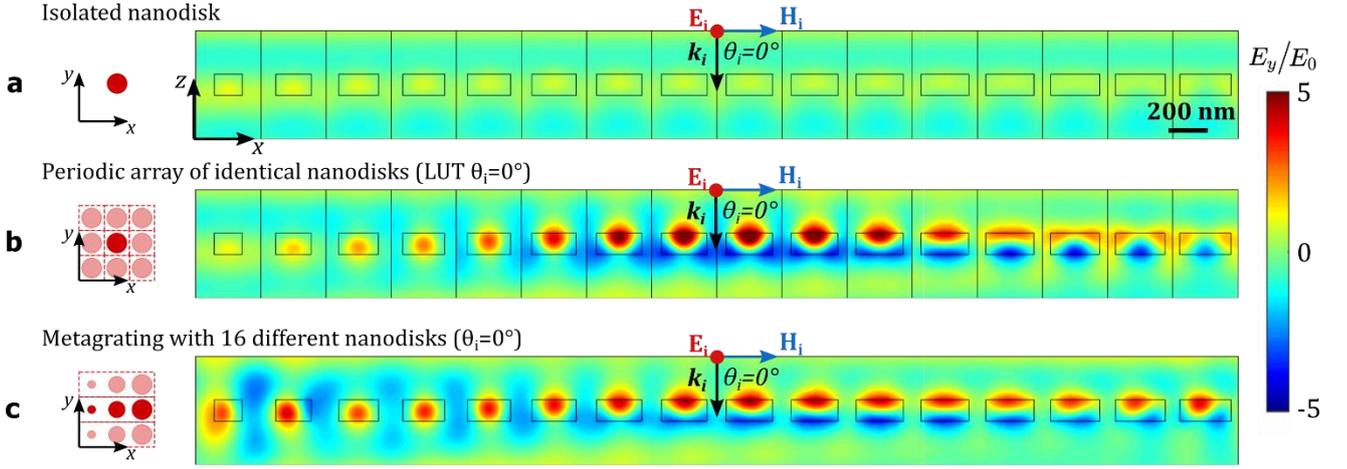

Figure 3. Analysis of the y-component of electric near-field maps in the xz plane of Huygens' meta-gratings. All the maps are obtained for the same polarization (TE) and wavelength ($\lambda = 1340$ nm). Incident electric/magnetic field orientations are reported with red/blue arrows. Field amplitudes are normalized to the incident field. **a** Isolated nanodisks. **b** Periodically arrayed nanodisks like in the lookup table (LUT) in Fig. 1b. The diameters are chosen to uniformly sample the $[0,2\pi]$ phase interval. **c** Meta-grating composed by associating 16 meta-atoms in a row. The first-order deviation angle is $\theta_B = 4.3°$. All the maps are obtained for normal incidence ($\theta_i = 0°$).

It is interesting to comparatively consider the near-field maps for the guidance approach, see Figs. 4a-b. Sixteen $TiO_2$ nanopillars are considered. They are periodically arrayed in 4a, exactly as for the computation of the lookup table of Fig. 1a. They form a $6.9\lambda$-period meta-grating in 4b. The grating efficiency is 92%. The two maps are very similar. In fact, the nanopillars behave almost independently, and thus, their near-field response weakly depends on the exact way they are arrayed [38].

The individual operation at the subwavelength scale has important consequences for light shaping: the nanowaveguides funnel the light vertically through the diffractive element, even if they are illuminated at oblique incidence [1,15]. Thus, the well-known shadowing effect that prevents conventional sawtooth diffractive elements from reaching high efficiencies for large deviations is circumvented. This quite unique property has been known since the 90's [38], see also [1,7] for a perspective discussion for optical imaging with high numerical aperture. It seems that the transient zone of the Huygens' nanodisks around the $2\pi$ phase jump just acts as a shadow zone, considerably reducing the efficiency of Huygens' meta-gratings especially for relatively small periods, see Fig. 2b.

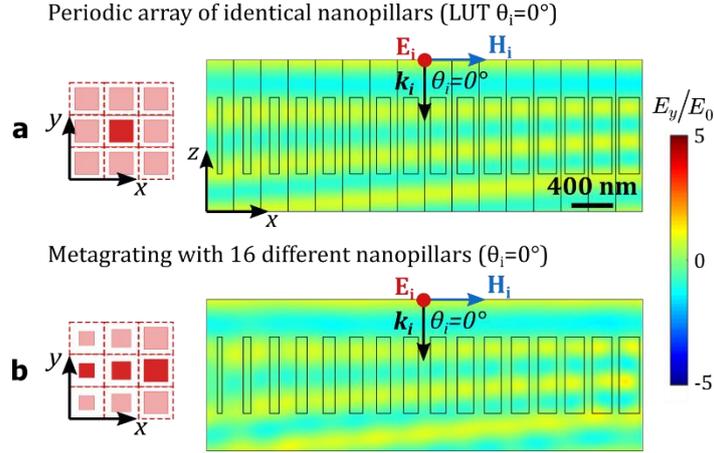

Figure 4. Analog of Figs. 3a and 3c for TiO$_2$-on-SiO$_2$ nanopillars. The $E_y$ maps are obtained for the same TE polarization and 633-nm wavelength. **a** Periodically-arrayed nanopillars like in the lookup table (LUT) of Fig. 1a. The pillar sizes are chosen to uniformly sample the [0,2π] phase interval. **b** Metagrating composed by associating the 16 nanopillars in a row. The first-order deviation angle is $\theta_B$=5.7°. Incident electric/magnetic field orientations are reported with red/blue arrows. The same color scale is used and the field amplitudes are normalized to the incident field.

### 4.2. Symmetry breaking: the hidden modes

After focusing our attention on the outermost 8 nanodisks around the phase jump, let us further consider the near-field maps of the 8 nanodisks of the central zone in Fig. 3c. Owing to the progressive variation in disk diameters (the diameter $d$ only changes by 3 nm on average between adjacent disks), one expects the maps to closely resemble the lookup-table maps in Fig. 3b. Although the resemblance is better than for the nanodisks in the transient zone, it is far from perfect – consider the field at the nanodisk rear and front interfaces – especially in view of the minute changes in diameter, $\Delta d/d \approx 5\%$, between adjacent pillars.

To explain the hypersensitivity of the metasurface to *gradual* and *tiny* variations, one may argue that the nanodisks behave as resonators, an effect that has been successfully employed in sensing [5]. Although reasonably valid, this argument alone does not appear sufficient. A symmetry argument should be added.

Symmetry breaking has surprising consequences in many fields, even with infinitesimally tiny modifications. In the present context, it is the discrete translational symmetry (i.e. periodicity) of the lookup-table geometries that is broken in meta-gratings.

To clarify the origin of the asymmetry, we have computed the Bloch modes for all the disk diameters of the lookup table for normal incidence. Up to six propagative modes are found for large disks, three being symmetric with respect to the vertical plane and the others being antisymmetric. Their effective indices (normalized propagation constant along the z-axis) are shown in Fig. 5a and their transverse field maps for a disk diameter of 550 nm are reported in the supplementary Fig. S3. Indeed, only the symmetric Bloch modes (blue curves in Fig. 5a) contribute to the lookup-table maps of Fig. 3b at $\theta_i = 0$. However, all modes, symmetric or antisymmetric, play an important role in meta-gratings, even for large periods implementing slowly varying phase gradients, and even at normal incidence.

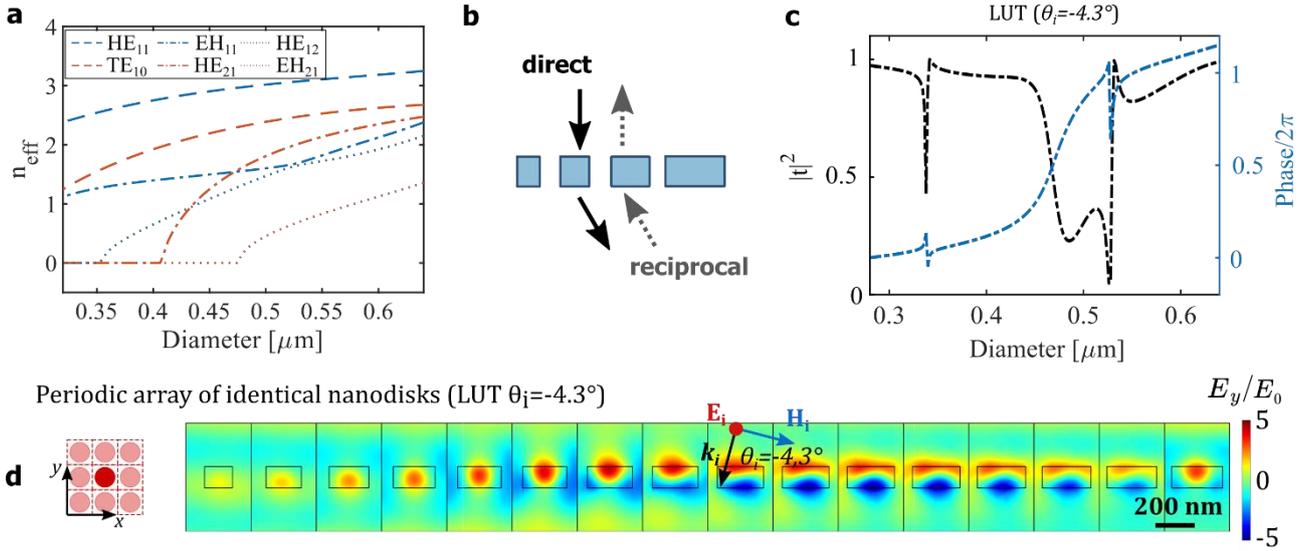

Figure 5. Reciprocity and hidden Bloch modes. **a** Effective indices of the three symmetric (blue) and three antisymmetric (red) non-evanescent Bloch modes of periodic arrays of identical Si nanodisks embedded in a host medium ($n_h = 1.66$) for $\Lambda = 666$ nm, $\lambda = 1340$ nm and $k_{//} = 0$. They are labelled according to the mode nomenclature of circular waveguide [39]. The field maps of the Bloch modes are plotted in the supplementary Fig. S3 for a diameter $d = 550$ nm. **b** Reciprocity argument to be considered for the design of a metasurface: any good design procedure should be valid for both the direct and reciprocal channels. **c** Look-up table for an incidence of $\theta_i = -4.3°$, which corresponds to the deviation angle into the first transmitted order of the meta-gratings in Fig. 3c. The difference with Fig. 1b obtained for $\theta_i = 0°$ is stringent. **d** Same as in Fig. 3b, except that $\theta_i = -4.3°$. Again, the difference is stringent. The Bloch modes are computed with the freeware RETICOLO [31].

To fully realize why that happens, let us consider a meta-grating designed to be blazed into the first transmitted order upon illumination by a normally incident plane wave (solid arrows in Fig. 5b). The same grating back-illuminated from the first transmitted order on the rear side (dashed arrows), will diffract light normally with the same efficiency. Now, owing to reciprocity, we realize that a proper design cannot rely on computational data involving symmetric Bloch modes only.

In Fig. 5c, we report the lookup table computed for an incidence angle $\theta_i = -4.3°$, which corresponds to the deviation angle of the meta-grating first order in Fig. 3c. We use the same axes as in Fig. 1b obtained for $\theta_i =$ 0° to allow for a direct comparison. The new transmittance displays deep dips, which are strongly polarization dependent (not shown), considerably reduce the efficiency and preclude a fine control of the imprinted phase. Finally, Fig. 5d shows the near-field maps corresponding to a lookup table computed for $\theta_i = -4.3°$. It is noticeable that the maps, albeit obtained for the same 16 nanodisks, strongly differ from those reported in Fig. 3b for $\theta_i = 0°$. Clearly the antisymmetric modes have a strong impact despite the small $\Delta k_{//}$ values. The essence of Huygens' flat-optics can be traced back to a series of pioneering works on individual *resonant* nanoparticles that scatter forward [20,21,36,37]. However, when they are arrayed, a collective resonance builds up and a strong spatial dispersion or grating-like behavior is implemented. Note that experimental demonstrations of Huygens metasurfaces sensitivity to incidence angle have been reported in [40].

Forward resonant scattering is successfully implemented in the lookup table of Fig. 1b with three Bloch modes, but this holds only for normal incidence under some symmetry restrictions. In fact, a smart design should harness a total of six modes for a broad range of geometrical parameters. Multimodal approaches are challenging for design in our opinion. In comparison, a single Bloch mode per polarization is involved in the operation of the $TiO_2$ nanowaveguides even at oblique incidence, owing to the choice of the periodicity that is forced to be smaller than the structural cutoff for which second-order propagative Bloch modes appear [14].

### 4.3. Lack of networkable Kerker-like resonant optical meta-atoms

Bloch modes are relevant for providing an insight into the properties of arrayed meta-atoms. In this last Subsection, we look back into our initial motivation of using nanotechnology to independently control rapidly varying phase (or amplitude, polarization) at the sub-wavelength scale. We thus consider the native isolated meta-atoms, which are at the root of the Bloch-mode properties previously analyzed. Again, by comparing the resonant and non-resonant approaches, we hope to grasp fundamental limitations.

Figure 6a shows the complex frequency plane of a single nanodisk exhibiting purely forward scattering at $\lambda = 1340$ nm, according to first Kerker condition (the magnetic and electric dipoles, $MD_x$ and $ED_y$, are equal, see Supplementary Section 4 for details). Not less than 8 (taking into account degeneracy) Mie-like modes are found in the quite narrow spectral region that we consider. These modes will be referred to as quasinormal modes (QNMs) to emphasize their non-Hermitian character [41]. The QNMs are normalized with the PML-normalization method using the QNMEig solver [42] of the freeware package MAN [43] based on COMSOL multiphysics. Four modes significantly contribute to the response for TE-polarized light: they are not pure multipoles, but referring to their dominant character, we may classify them as two magnetic dipoles (MDx and MDz), one electric dipole (EDy) and one electric quadrupole (EQ).

When the incident plane wave impinges normally on the nanodisk ($\theta_i = 0$ and $\varphi = 0$), only $MD_x$ and $ED_y$, which are symmetric with respect to the $x$ axis, are excited. The antisymmetric modes, $MD_z$ and EQ, contribute only for $\theta_i \neq 0$. They may even dominate the response, as shown in Fig. 6b, where the angular dependence of the excitation efficiency $|\alpha|^2$ of each QNM is plotted for an incident plane wave polarized along $y$ (see [44] for details on the definition and computation of $\alpha$). The disruptive role of $MD_z$ and EQ in the electromagnetic coupling with neighbors for beam shaping is more clearly evidenced with the far field patterns shown in Fig. 6c. Besides the fact that all radiation diagrams exhibit forward scattering, two interesting points are noticeable. First, even at normal incidence (yellow curve), the horizontally scattered field is not null, implying that a significant long-range interaction is present in any array. Second, as $\theta_i$ increases, the $MD_z$ and EQ contributions become significant and the in-plane scattering becomes huge, see the black curve obtained for $\theta_i = 40°$.

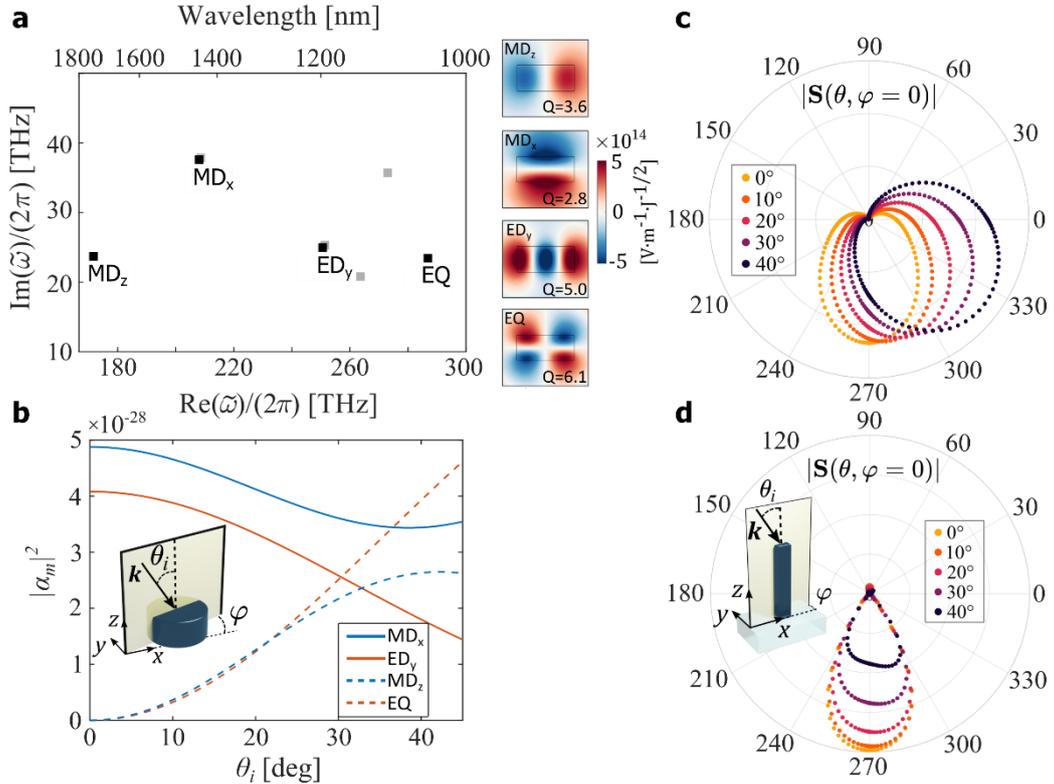

Figure 6. Difficulty to implement smart meta-atoms is highlighted by a study of the scattering properties of an isolated Si nanodisk with radius $r = 243$ nm and height $h = 220$ nm embedded in a host medium ($n_h = 1.66$). **a** Complex frequency plane locating the Mie resonances. The dominant ones are shown with dark squares. Inset: quality factors ($Q = \text{Re}[\tilde{\omega}]/2\text{Im}[\tilde{\omega}]$) and near-field maps of $E_y$. The modes are normalized with the PML-normalization method and computed with the freeware package MAN [42-43] . The normalization allows for a direct visual comparison of the interaction strength of the mode with any driving field. **b** Square modulus of the excitation coefficients vs. incident angle $\theta_i$ for the four dominant resonances. The incident plane wave is polarized along $y$. **c** Far-field radiation diagram in the $xz$ plane for various incident angles $\theta_i$. **d** Same as in (c) for a TiO$_2$/SiO$_2$ nano-waveguide (size 147 nm). The radiation diagrams are computed with the near-to-far field transformation package RETOP [45].

For the nanowaveguide approach, the coupling is essentially short range, through the evanescent field of the nanopillar guided modes, as shown in Fig. 6d which conveys the idea that long-range in-plane electromagnetic interaction is largely removed with the guidance approach. The arrayed nanopillars operate nearly independently of each other [38].

For a single polarization, the nanodisk response is dominantly driven by 4 resonances, which result in 6 non-evanescent Bloch modes in the arrayed picture. As shown with the reciprocity argument, all these resonances are necessarily excited in graded metasurfaces. This gives rise to non-local behaviors that are substantially sensitive to periodicity breaking and result in a strong spatial dispersion exemplified in Fig. 5c. To avoid the formation of collective resonances, long-range in-plane interactions have to be minimized and therefore, one is inclined to design meta-atoms *solely* supporting equally-strong MDx or EDy resonances (assuming $y$-polarization): all the other resonances, MD$_z$, ED$_z$ and higher order ones which necessarily exist, should be pushed away in the complex plane, out of the frequency window of interest.

The design of meta-atoms offering a prescribed phase and amplitude relation between two dipole moments and 'isolation' in the complex plane is a truly difficult task. Recently, a related study of similar complexity has been undertaken by some of the authors for designing a Janus nanoparticle [46] operating over broad spectral and angular ranges. Despite the use of advanced numerical tools, success has been mitigated owing to a spurious QNM that could not be eliminated, restricting the operation of the designed nanoparticle to a single polarization for which the extra QNM is not excited [47]. More generally, the design issue is reminiscent of the remarkable work performed during two decennia towards the implementation of 3D negative-index optical materials. Even with tiny plasmonic meta-atoms that are significantly smaller than the wavelength [48], isotropic negative-index materials have somehow never been realized, not at any visible wavelength, not even in the microwave region. Strong spatial dispersion could not be avoided [49-52].

The difficulty is exacerbated with dielectric meta-atoms, for which artificial magnetism requires resonators with a size that is typically equal to the wavelength in the particle. Dielectric meta-atoms are thus multipolar in nature and isolation is hard to achieve. Please note, for example, that the design of centro-symmetric nanoparticles is not suitable, since the prescribed $MD_x$ and $ED_y$ resonances would automatically result in $MD_z$ and $ED_z$ resonances. We are facing fundamental limitations [52].

To further support our analysis, finally note that excellent performance is achieved with dielectric meta-atoms for ultra-flat *reflecting* metasurfaces operating with a single electric-dipolar resonance only [53]. Purely electrical dipolar resonances are simple to isolate in the complex frequency plane. The challenge arises when trying to design sophisticated resonances.

## 5. Conclusion

The comparison of the principles of operation of nanowaveguide phased-array antennas and ultra-flat Huygens' metasurfaces shows that the former remains the simpler and more efficient way to imprint rapidly varying phases at optical frequencies. Provided that the waveguide spacing is suitably chosen [1], neither too small to avoid the homogeneization regime, nor too large to remain below the structural cutoff for which second-order progressive Bloch modes appear, good performance is obtained at the design frequency.

It is difficult to design resonant (and thus flat) meta-atoms that offer sophisticated and isolated resonances at optical wavelengths with dielectric materials. Although alternative designs maintaining all the benefits of ultra-flat nanostructured elements are not strictly ruled out by our analysis, we believe that the fundamental limitations we bring up in Section 4 should be considered before tackling the design challenge. We wish that pointing out the different facets of the problem will help circumvent the current limitations.

**Supporting Information**

Supporting Information is available from the Wiley Online Library or from the author.

**Funding sources and acknowledgments**

PL and GL acknowledge NOMOS Project (ANR-18CE24-0026) for financial support. MLB and QL are funded by individual investigator grants from the department of energy (Grant DE-FG07-ER46426) and Airforce (Grant FA9550-18-1-0323). The authors acknowledge the challenging and valuable advises of four anonymous reviewers who helped deepening the analysis.

# Supplementary Information

# Fundamental limitations of Huygens' metasurfaces for optical beam shaping

*Carlo Gigli[1], Qitong Li[2], Pierre Chavel[3], Giuseppe Leo[1], Mark Brongersma[2], Philippe Lalanne[4*]*

[1] *Matériaux et Phénomènes Quantiques, Université de Paris & CNRS, 10 rue A. Domon et L. Duquet, 75013 Paris, France*
[2] *Department of Materials Science and Engineering, and Geballe Laboratory for Advanced Materials, Stanford University, Stanford, California 94305, USA*
[3] *Laboratoire Hubert Curien, Université de Lyon, Univ. Jean Monnet de Saint Etienne, Institut d'Optique Graduate School, CNRS UMR 5516, 42000 Saint Etienne, France*
[4] *Laboratoire Photonique, Numérique et Nanosciences (LP2N), Institut d'Optique Graduate School, Univ. Bordeaux, CNRS, 33400 Talence Cedex, France.*
*\*Corresponding author:* Philippe.lalanne@institutoptique.fr

## Contents

## S1. Multipolar Analysis

The well-known near-unity transmittance for Kerker metasurfaces is achieved when the meta-atoms in a uniform array are engineered to satisfy the first Kerker condition, where the radiation from symmetric multipoles perfectly destructively interfere with that from anti-symmetric multipoles in the backscattering direction [1]. However, it is questionable that whether the Kerker condition can still be held for the same-sized meta-atoms in a non-uniform array, as the coupling between meta-atoms is not negligible. In order to answer this question quantitatively, we perform multipolar analysis for different-sized meta-atoms in a meta-grating as well as in a uniform array. The polarization vector **P** inside each meta-atom volume $\Omega$ is expanded in multipole moment contributions in Cartesian coordinates according to [2]:

$$\mathbf{p} = \int_\Omega \mathbf{P}(\mathbf{r'})d\mathbf{r'}$$

$$\mathbf{m} = -\frac{i\omega}{2}\int_\Omega [\mathbf{r'} \times \mathbf{P}(\mathbf{r'})]d\mathbf{r'}$$

$$Q = 3\int_\Omega [\mathbf{r'}\mathbf{P}(\mathbf{r'}) + \mathbf{P}(\mathbf{r'})\mathbf{r'}]d\mathbf{r'}$$

$$M = -\frac{2i\omega}{3}\int_\Omega [\mathbf{r'} \times \mathbf{P}(\mathbf{r'})]\mathbf{r'}d\mathbf{r'}$$

$$\mathbf{T} = \int_\Omega \{2\mathbf{r'}^2\mathbf{P}(\mathbf{r'}) - [\mathbf{r'} \cdot \mathbf{P}(\mathbf{r'})]\mathbf{r'}\}d\mathbf{r'}$$

where **p**, **m**, and **T** represent electric, magnetic, and toroidal dipole vectors while $Q$ and $M$ are electric and magnetic quadrupole tensors. We note that electric dipole, toroidal dipole, as well as magnetic quadrupole have symmetric radiation patterns, while magnetic dipole and electric quadrupole are anti-symmetric moments. By summing up the back-scattered electric fields in normal direction from symmetric moments (noted as $E_{sca\_sym}$) and anti-symmetric moments (noted as $E_{sca\_antisym}$) separately, we can evaluate the Kerker condition as summarized in Fig. S1. Circle (resp. star) marks represent the scattered electric fields from individual meta-atoms in a meta-grating (resp. uniform array), and the mark color (from red to blue) represents the different-sized meta-atoms (from small to large).

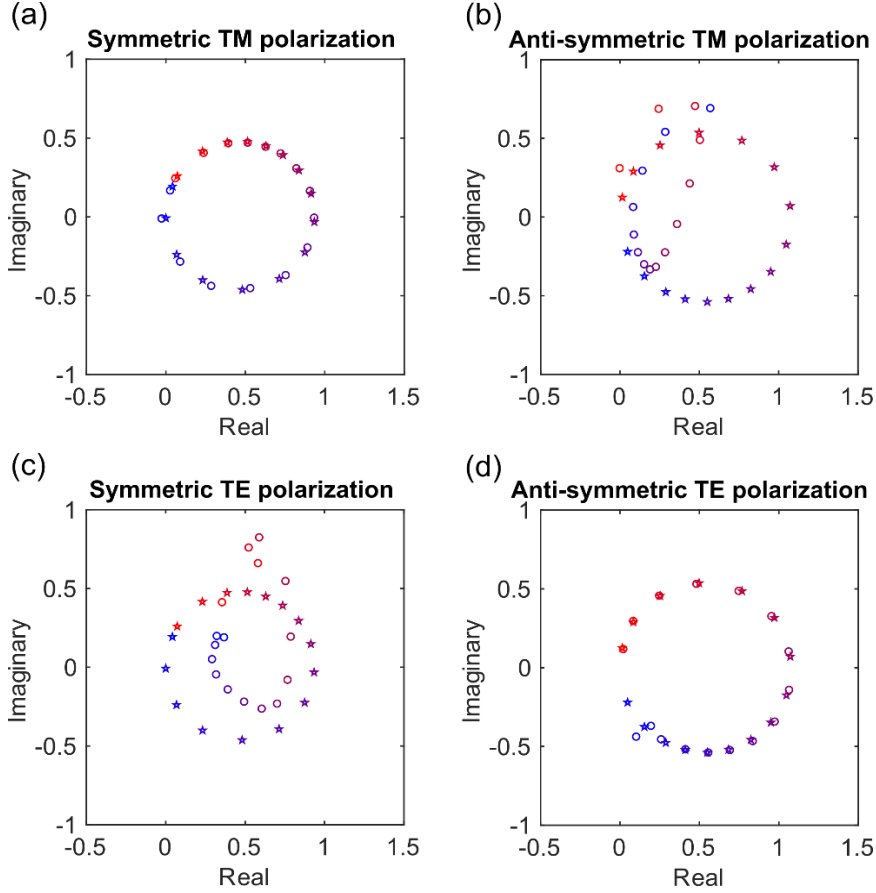

**Fig. S1**. Multipolar decomposition of electromagnetic near field in silicon nanodisks metasurfaces. (a) The scattered electric fields from symmetric and (b) anti-symmetric multipoles for meta-atoms in a meta-grating (circle marks) and in a uniform array (star marks) under TM polarization. (c, d) Same for TE polarization. The shading from red to blue denotes the increasing nanodisk size: red stands for the smallest ($r = 143$ nm) and blue for the largest ($r = 261$ nm) nanodisks.

We firstly find that $E_{sca\_sym}$ overlaps with $E_{sca\_antisym}$ nicely for meta-atoms in a uniform array (star marks), as they form similar circular loops in Figs. S1a and d. This indicates that the Kerker condition is very well satisfied for all these different-sized meta-atoms in a uniform array. However, Figure S1b shows that when the meta-atoms are arranged in a meta-grating (circle marks), $E_{sca\_antisym}$ under TM polarization significantly deviates from the original circular loop. Similar deviations are also found for $E_{sca\_sym}$ under TE polarization in a meta-grating, see Fig. S1c. As a result, for meta-atoms in a meta-grating, the back scattered electric fields from symmetric and anti-symmetric multipoles do not match for both polarizations, and therefore the original Kerker condition achieved in a uniform array is largely destroyed. For this reason, we expect that the backscattering from meta-atoms should become considerable when they are arranged in a non-uniform way in order to generate phase masks, though these meta-atoms do meet Kerker conditions in uniform arrays. We emphasize that the observed large deviation does not come from rapid phase variation, as the analyzed meta-grating is composed of super cells containing sixteen meta-atoms in total (blazing angle 7.2°). We believe that the above multipolar analysis quantitatively illustrates why the real diffraction efficiency of a Kerker metasurface is much lower than the calculated lookup table.

## S2. Huygens' meta-gratings – TM polarization

In the main text, by comparing the near-field maps in Fig. 3b (periodic array, lookup table) and 3d (meta-grating supercell composed by associating 16 nanodisks of different size), we observed a significant difference, especially for the nanodisks on the right and left sides of the period close to the $2\pi$ phase jump, and concluded that the relatively poor performance of Huygens' meta-gratings is due to shadowing. The analysis was performed for TE polarization.

In Fig. S2, we provide the same comparison for TM polarization. Exactly the same conclusion holds.

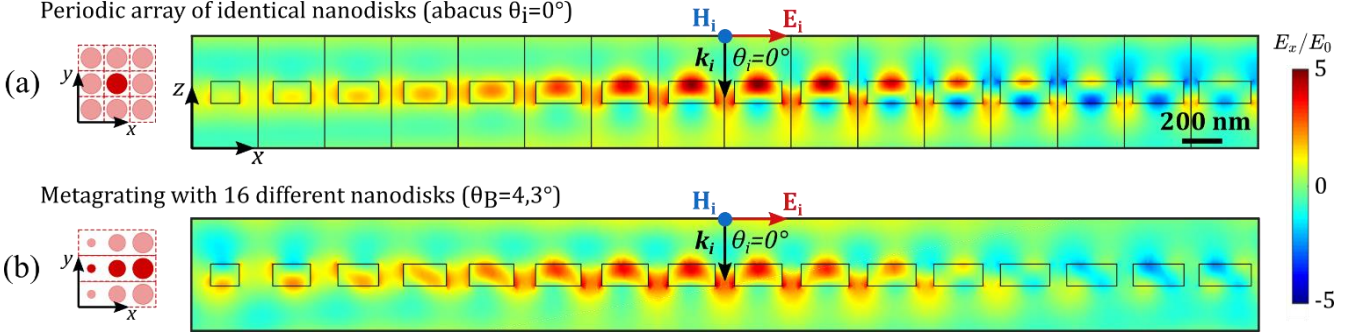

**Fig. S2**. Same as in Fig. 3(b) and 3(d) in the main text for TM polarization. Near field maps of Huygens' blazed grating operated under TM polarization. Top: near-field distribution of Re($E_x$) for 16 arrayed nanodisks providing a linear gradual phase variation from 0 to $2\pi$ placed. Bottom: comparative analysis for the same sixteen nanodisks in a meta-grating supercell composed by associating the 16 nanodisks.

## S3. Bloch Modes analysis

In this section, we estimate the number and nature of Bloch modes that are involved in the two transmission of the lookup tables of Fig. 3(b) and 3(c), focusing our attention on the contribution of non-evanescent Bloch modes. In our view, the embodiment of many Bloch mode quantifies the richness of the Mie resonance. It also quantifies the number of degrees of freedom the designer should implicitly take into consideration for successful operation.

Using the freeware Reticolo [3], we first compute the normalized propagation constant (or effective index $n_{\text{eff}}$) of the propagative Bloch modes (with $Im(n_{\text{eff}}) = 0$) in periodic arrays of nanodisks as a function of the disk diameter, see Fig. 4. Two incidence angles, $\theta_i = 0$ and $-4.3°$, are considered.

Up to six non-evanescent Bloch modes are found for large resonators. The near-field distributions of the Bloch modes computed for a nanodisk diameter $d = 550$ nm excited are reported in Fig. S3(c). We label them according to the nomenclature of the modes of a circular waveguide [4]. It is noteworthy that half of them are symmetric (S1-3) and half antisymmetric (A1-3). The antisymmetric modes are not excited at normal incidence; the symmetry breaking between $\theta_i = 0$ and $-4.3°$ is evidenced by the anti-crossing in the curve $n_{\text{eff}}(d)$ between S2 and A2 for an incidence angle $\theta_i = -4.3°$.

Then the transmission coefficient is reconstructed for each diameter considering an increasing amount of propagative Bloch modes. At normal incidence, we just consider the 3 symmetric modes and accurately reconstruct the transmission coefficient computed with RCWA. Numerical details explaining how to reconstruct scattering coefficients with a subset of Bloch modes can be found in [5]. For an incident angle $\theta_i = -4.3°$, the contribution of antisymmetric modes becomes relevant, as evidenced by the remarkable drop in the transmittivity. Please note that, in contrast with the normal incidence case, the reconstruction with the 6 non-evanescent Bloch modes is only qualitatively accurate, and a discrepancy remains between the solid-black (RCWA) and dotted-orange (6-mode approximation) curves. This evidences the embodiment of additional evanescent modes that funnel energy through the metasurface. This mechanism is favored for small $Im(n_{\text{eff}})$ values and for small grating thicknesses, as it is the case for Huygens' metasurfaces [6].

This Bloch mode analysis highlights the multimodal nature of Huygens metasurfaces; many degrees of freedom and physical effects have to be carefully engineered for successful design.

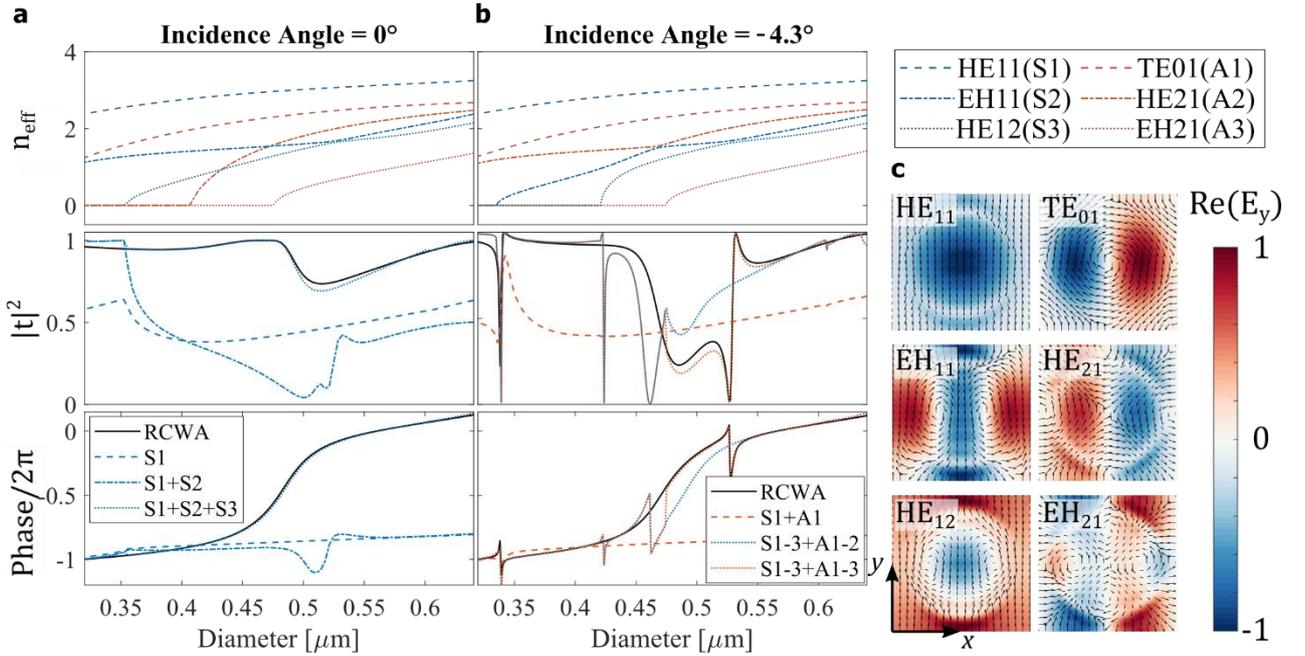

**Fig. S3**. a-b Bloch modes analysis of lookup table calculations. Top: Diameter-dependence of the effective indices of the six propagative Bloch modes for two incident angles $\theta_i = 0°$ and $-4.3°$, corresponding to the lookup table shown in Figs. 3b and 3c, respectively. Bottom: Transmission coefficient reconstruction (intensity and phase) with an increasing number of Bloch modes (colored lines). The full-wave RCWA results are shown with the solid-black curve. c Near-field distribution of $Re(E_y)$ for the six propagative modes for a nanodisk with a diameter $d = 550$ nm.

## S4. Isolated Huygens' source

The geometry of an isolated Huygens' source was optimized analyzing the multipolar contribution to the scattering from a Si nanodisk with height $h = 220$ nm embedded in a uniform medium with index $n_b = 1.66$ at fixed excitation wavelength $\lambda = 1340$ nm. The near-field was computed in COMSOL Multiphysics, and the multipolar contributions to the scattering, see Fig. S4a, were obtained according to the formulations in [7]. For a radius $r = 243$ nm, the electric dipole along y and the magnetic dipole along x, computed as in [8], have almost same amplitude and phase confirming the fulfillment of first Kerker condition.

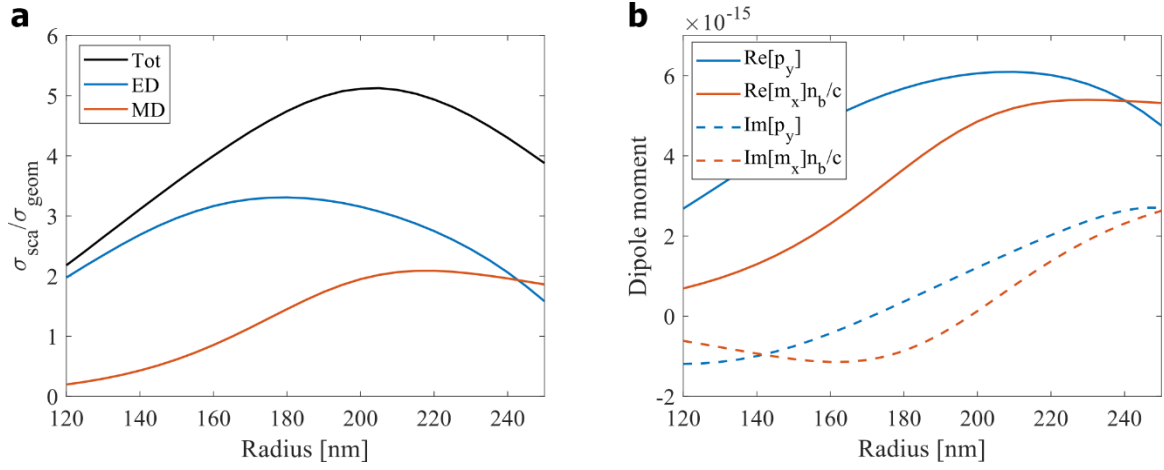

**Fig. S4.** Scattering from an isolated Si nanodisk with height 220 nm. **a** Total (black), electric (blue) and magnetic (red) dipole contributions to scattering cross section normalized by geometrical one vs. nanodisk radius. **b** Electric (blue) and magnetic (red) dipole moments in real (solid) and imaginary (dashed) parts vs. nanodisk radius.